\documentclass[%
 reprint,
nofootinbib,
amsmath,
amssymb,
aps,
]{revtex4-2}

\usepackage{graphicx}% Include figure files
\usepackage{dcolumn}% Align table columns on decimal point
\usepackage{bm}% bold math
\usepackage{braket}
\usepackage[hidelinks]{hyperref}
\usepackage{relsize}
\begin{document}

\preprint{APS/123-QED}

\title{Nanoparticle Interferometer by Throw and Catch}

\author{Jakub Wardak $^{1}$}
\author{Tiberius Georgescu $^{1}$}
\author{Giulio Gasbarri $^{2}$}
\author{Alessio Belenchia $^{3, 4}$}
\author{Hendrik Ulbricht $^{1,}$}
\email{Correspondence: h.ulbricht@soton.ac.uk}
\affiliation{
$^{1}$ School of Physics and Astronomy, University of Southampton, SO17 1BJ, Southampton, United Kingdom\\
$^{2}$ Física Teòrica: Informació i Fenòmens Quàntics, Department de Física, Universitat Autònoma de Barcelona, 08193 Bellaterra (Barcelona), Spain\\
$^{3}$ Institut für Theoretische Physik, Eberhard-Karls-Universität Tübingen, 72076 Tübingen, Germany \\
$^{4}$ Centre for Theoretical Atomic, Molecular and Optical Physics, School of Mathematics and Physics, Queen's University Belfast, Belfast BT7 1NN, United Kingdom}

\date{\today}% It is always \today, today,
             %  but any date may be explicitly specified

\begin{abstract}
Matter wave interferometry with increasingly larger masses could pave the way to understanding the nature of wavefunction collapse, the quantum to classical transition, or even how an object in a spatial superposition interacts with its gravitational field. In order to improve upon the current mass record, it is necessary to move into the nanoparticle regime. In this paper, we provide a design for a nanoparticle Talbot--Lau matter wave interferometer that circumvents the practical challenges of previously proposed designs. We present numerical estimates
%simulations 
 of the expected fringe patterns that such an interferometer would produce, considering all major sources of decoherence. We discuss the practical challenges involved in building such an experiment, as well as some preliminary experimental results to illustrate the proposed measurement scheme. We show that such a design is suitable for seeing interference fringes with $10^6$~amu SiO$_2$ particles and that this design can be extended to even $10^8$~amu particles by using flight times below the typical Talbot time of the system.
\end{abstract}

%\keywords{Suggested keywords}%Use showkeys class option if keyword
                              %display desired
\maketitle

%\tableofcontents

\section{Introduction}

Interferometry techniques have historically been the most-popular method for demonstrating wave-like behaviour. Perhaps the most-famous example of this was Young’s double slit experiment, performed in 1801, which demonstrated the wave nature of light \cite{youngdoubleslit}. The subsequent works of Planck and Einstein showed that light also possesses particle-like features in the form of photons. The idea of wave--particle duality, extended to also encompass matter, was then famously formalised by Louis de Broglie in his 1924 Ph.D. thesis, where he claimed electrons could also behave like waves with their wavelength being dependent on their momentum~\cite{de1925research}. His prediction was later verified by Davisson and Germer in 1927, who, like Young, performed a double slit experiment to show interference in electrons~\cite{davisson1928reflection}. 

Since wave--particle duality is a core tenant of quantum mechanics, testing this phenomenon at increasingly larger scales could give us some much-needed insight into poorly understood areas of quantum mechanics such as the nature of wave function collapse, the so-called ‘measurement problem’~\cite{RevModPhys.75.715,RevModPhys.76.1267}, or how massive quantum objects interact with their gravitational fields~\cite{PhysRevLett.119.240401,PhysRevLett.119.240402}. Much progress has been made in the field of matter wave interferometry since the electron double slit experiment, initially with atoms of increasing mass~\cite{atominterf,gould1986diffraction,keith1988diffraction,borde1989atomic,kasevich1991atomic} and, then, with larger and larger macro-molecules~\cite{hornberger2012colloquium}. The current mass record for observing matter wave interference is held by Markus Arndt’s group in Vienna, who used a Talbot--Lau interferometer (TLI) scheme to demonstrate the matter wave interference of molecules of masses up to 25,000 atomic mass units (amu)~\cite{molleculeinterf}. These near-field techniques were pioneered by Clauser and Li \cite{clauser}, who demonstrated the Talbot--Lau interferometry in atoms before the rise of cold atomic ensemble experiments enabled by laser cooling techniques. These techniques were then famously extended to higher mass matter wave interferometry in 1999, where the interference patterns of buckminsterfullerene (C$_{60}$) were shown~\cite{c60} in the far-field and shortly later in the molecular TLI scheme as well~\cite{brezger2002matter}.

Further probing of the fundamentals of quantum mechanics will require larger mass matter wave interferometric experiments~\cite{arndt2014testing}. However, the molecular techniques used in these experiments have inherent problems with scaling up to ever larger masses. Especially challenging is the provision of coherent particle beam sources, given there are limited capabilities for cooling the centre of mass motion of large molecules and clusters~\cite{juffmann2013experimental}. For this reason, in this paper, we aimed to introduce a new scheme for the detection of interference effects using SiO$_{2}$ nanoparticles using a ‘throw and catch’ design. Silica nanoparticles have large polarisabilities, which opens the opportunity for manipulation and, indeed, cooling to the motional ground state by optical techniques beside others in the emerging field of levitated mechanics~\cite{gonzalez2021levitodynamics}. The interferometer approach will be heavily based on the proposal suggested by Bateman et al. in 2014 \cite{bateman2014near}, with the throw and catch scheme intended to alleviate the practical issues of the original proposal, such as inefficient reloading. We will show the expected interference patterns produced whilst using realistic experimental parameters for both $10^6$~amu and $10^8$~amu particles including all major sources of decoherence. 

\section{Experimental Set-Up}

The key idea behind this proposal is to use the core concepts presented in \mbox{Bateman et al.} whilst implementing a method of reusing the same particle throughout the generation of the interference pattern. This will alleviate practical issues of loading and re-using nanoparticles since re-trapping an appropriate particle can take hours. It will also give us a completely identical source for each run of the experiment, thereby reducing decoherence effects from non-identical sources. This is demonstrated in Appendix A Figure~\ref{fig:A1}.

\begin{figure}[b]
\includegraphics[width=8cm]{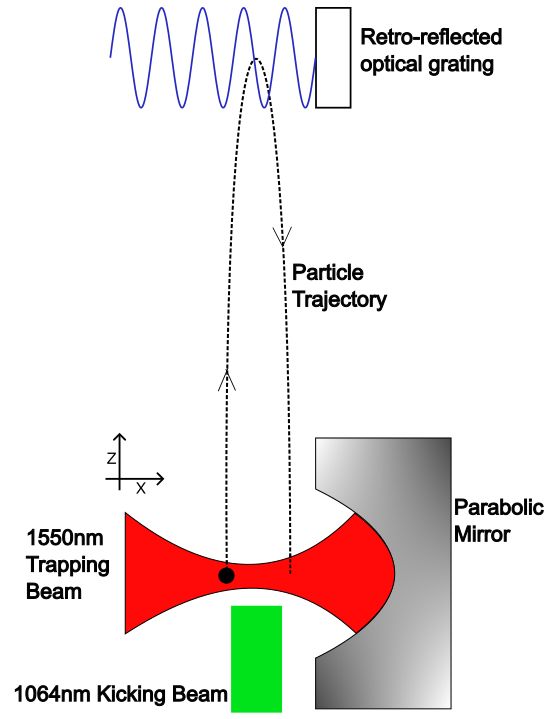}
\centering
\caption{\label{fig1} Schematic %MDPI: 1. We moved it here after first citation, same below, please confirm. 2. please add explanation of different colors.
 of the proposed experimental setup, with different coloured beams representing different wavelength lasers. The particle is shown as being kicked when offset from the trapping centre to demonstrate the generally parabolic trajectory. Please note that, in the following text, the parabolic mirror will be assumed to be placed such that the longitudinal axis of the trapping light is along the $z$ direction. The kicking pulse will then be focused by the parabola onto the particle. This figure aims to give a rough concept of the design.}
\end{figure}

A rough outline of the scheme is shown in Figure~\ref{fig1}. We begin with trapping a silicon dioxide nanosphere using a parabolic trap detailed in %MDPI: There is no ref [23] between ref [22] and [24], please add cite of ref [23]. % ref [23] is referrenced in the end notes only.
 \cite{parabloictrap}. We used a 1550 nm laser with an incident power of approximately 100 mW for the purposes of trapping using the standard optical tweezer technique. The light scattered by the particle will be collimated by the parabola and detected to track the particle motion. The particle behaves as a simple harmonic oscillator, and its motion modulates the phase of the trapping laser beam. By interfering the scattered light with the light that misses the particle, we can extract this phase information and identify the particle's modes of motion. We then used this information to implement feedback cooling, via a lock-in amplifier, which modulates the trapping light, to decrease the motional temperature of the particle to approximately 1 mK along the direction of the laser grating. This results in a momentum uncertainty $\sigma_p/m$ of less than $0.1$~cms$^{-1}$ and a positional uncertainty of less than $1$~nm for a $10^8$~amu particle. The particle in our trap, hence, acts as a coherent source of matter waves \footnote{Interference %MDPI: footnote is not allowed in this journal, we changed it to note, please confirm.
 effects between the coherent part of the wave function can only occur if the size of the original source is smaller than the grating period $\sigma_x /d < 1$. One furthermore needs $\sigma_p d/h \gg 1$, to ensure that the initial trapped state extends over many grating momenta, a necessary condition to guarantee the validity of the theoretical model used to describe the interferometric setup~\cite{bateman2014near,nimmrichter2008theory}. Both of these conditions are fulfilled for the case of study presented here.}.
 
 After the cooling, we turned off the trapping light and applied a vertical impulsive force via the action of a 1064~nm laser pulse. The power of the laser was tuned to achieve the required flight time to generate, via the Talbot effect, an experimentally detectable quantum interference pattern. The Talbot effect is a near-field interference effect, where the image of the diffraction grating, that a wave passes through, is repeated at regular intervals. The distance between these intervals is known as the Talbot length, and by knowing the velocity of our particle, we can also think of this as a Talbot time. In order to be confident of seeing such an interference pattern, the flight time of our particle ought to be on the scale of the Talbot time, which is around $2.8$~s, for the case of a $10^8$~amu particle. However, we will show that this is not such a strict requirement. Halfway through the particle's total flight time, at the peak of its trajectory, a secondary laser pulse of wavelength $213$~nm, and nanosecond duration, was imparted onto the particle. This is the grating pulse. This pulse was retro-reflected by a mirror in order to form an optical grating through which the matter wave is diffracted. This process introduced a position dependent phase shift to the wave function of the matter wave, hence forming the basis of our interference pattern.

 The final stage for our proposed interferometer was the measurement and ‘catch’. The optical trap was turned back on once the particle was just above the height it started at in order to slow down and recapture it. Precise timing will be important here, since, if the trap is turned on too early, the initial re-entry will accelerate the particle so much that it will fall out of the trap. This will be explained in more detail below. Once the particle is recaptured, we will once again be able to track its position as discussed earlier. By extrapolating this information back to the time at which the particle entered the trapping centre, we can measure the position it landed at, as demonstrated in Hebestreit et al. in 2018~\cite{measurementconcept}. We will require roughly a $10$~nm precision in order to detect our interference patterns. Initial testing showed that this should be achievable even with rudimentary data extraction techniques, with work ongoing on more advanced methods. The particle was then re-cooled and prepared to its initial state as described above, which only required a few trapping cycles and, hence, does not introduce major heating effects. Over many runs, a probability distribution of particle arrival locations was produced, which can be compared to the expected classical ballistic and quantum wave-like patterns discussed in the following.

\section{Theoretical Model} 

\subsection{Background}

As mentioned before, the design of the throw and catch experiment maintains the core aspects of the proposal set out in Bateman et al. \cite{bateman2014near}. This allowed us to work with the existing theory presented in Bateman et al. and, later, built upon in Belenchia et al. \cite{belenchia}. The only differences that need to be considered for the throw and catch variant are some additional decoherence effects, which will be discussed shortly. For this reason, we only give a brief overview of the theoretical background here, deferring to the aforementioned publications for a more-detailed description. 

We begin our experiment with the nanoparticle trapped by our $1550$~nm optical trap and behaving as a harmonic oscillator. This state is well described as a Gaussian thermal state of motion with standard deviations in position and momentum given by $\sigma_x = \sqrt{k_B T/4 \pi^2 m \nu^2_M}$ and $\sigma_p = \sqrt{m k_B T}$, respectively. Here, $T$ is the motional temperature of the oscillator, $m$ is the mass of our particle, and $\nu_M$ is the trap frequency as defined from the standard harmonic oscillator relationship between restoring force and position. At this point, it is important to note that our grating is a uniform standing wave oriented along the $x$-axis, as shown in Figure~\ref{fig1}. Then, assuming the laser waist in the $y$ direction is large enough and that the overall $y$ displacement is small throughout the journey, the grating should have a negligible impact on the evolution of the state in the $y$ and $z$ directions. This allowed us to effectively neglect these directions when evaluating our final interference pattern since this pattern will only exist along the $x$ direction. This is a crucial similarity between our work here and the preceding work in Bateman et al. \cite{bateman2014near}. 

We begin with the free evolution of the particle, after the kick, up to its peak. This occurs in a time $t_1$. Then, the interaction with the grating is followed by another period of free evolution as the particle falls back down into the trap, where its position is finally measured. This second period of free evolution occurs in a time $t_2$, and in our case, $t_1 = t_2$. We assumed the kick procedure can be modelled as a positive shift of the vertical velocity distribution of the particle state. Under this assumption we can use the expression derived in~\cite{bateman2014near}  to describe the final probability density function, upon measurement, that is:
% \begin{equation}
% \begin{aligned}
% w(x) & =\frac{m}{\sqrt{2 \pi} \sigma_p\left(t_1+t_2\right)} \sum_n \left\{ B_n\left(\frac{n t_1 t_2}{t_{\mathrm{T}}\left(t_1+t_2\right)}\right) \\
% & \times \exp \left[\frac{2 \pi i n x}{D}-\frac{2 \pi^2 n^2 \sigma_x^2 t_2^2}{d^2\left(t_1+t_2\right)^2}\right] \right\}.
% \end{aligned}\label{eq:interference}
% \end{equation}

\begin{equation}
\begin{aligned}
w(x) & =\frac{m}{\sqrt{2 \pi} \sigma_p\left(t_1+t_2\right)} \mathlarger{\mathlarger{\sum_n}} \Biggl\{B_n\left(\frac{n t_1 t_2}{t_{\mathrm{T}}\left(t_1+t_2\right)}\right) \\
& \times \exp \left[\frac{2 \pi i n x}{D}-\frac{2 \pi^2 n^2 \sigma_x^2 t_2^2}{d^2\left(t_1+t_2\right)^2}\right]\Biggr\}.
\end{aligned}\label{eq:interference}
\end{equation}

This expression is of the form of a periodic fringe pattern oscillating in the period $D = d(t_1 + t_2)/t_1$, with $d$ being the grating spacing given by $\lambda_G = 2d$. $t_T$ in the above expression is the Talbot time given by $t_T = m d^2 /h$, which sets the time scale that our nanosphere wave packet needs to evolve for in order to reasonably expect to see a near-field interference pattern. Finally, the $B_n$ terms are the so-called Talbot coefficients, which characterise the general shape of the interference pattern. For nanoparticles in the Rayleigh limit, that is whose linear dimension is sufficiently smaller than the grating period, these coefficients take the form~\cite{nimmrichter2014macroscopic}:
\begin{equation}\label{eq:talbot}
B_n(u) = J_n(\phi_0 \sin(\pi u)),
\end{equation}
where $J_n$ denotes a Bessel function of the first kind. The $\phi_0$ in the above expression is referred to as the phase modulation parameter and comes from the expression for the effect that the optical grating has on the particle's wave function. In the longitudinal eikonal approximation~\footnote{The validity of this approximation requires the interaction time between the system and the optical grating to be negligible with respect to the characteristic time of the free system's dynamics~\cite{nimmrichter2014macroscopic,nimmrichter2008theory}.} and ignoring incoherent effects, the quantum state evolves as:
\begin{equation}
\langle x | \psi\rangle \rightarrow \exp \left(i \phi_0 \cos ^2 \left( \frac{\pi x}{d} \right) \right)\langle x | \psi\rangle,
\end{equation}
where
\begin{equation}\label{eq:phi0}
\phi_0 = \frac{2 \operatorname{Re}(\alpha) E_G}{\hbar c \epsilon_0 a_G}.
\end{equation}
Here, %MDPI: please confimr if keep all noindent in this manuscript.
 $\alpha$ denotes the particle's static polarizability, whilst $E_G$ and $a_G$ are the energy and spot size of the optical grating pulse, respectively.

{It should be noted that, also, classical particles traversing the optical grating, and moving on ballistic trajectories, give rise to a shadow fringe pattern~\cite{hornberger2012colloquium,bateman2014near,belenchia}. This classical pattern, in the Rayleigh limit \footnote{For the case of finite-size particles, we refer the reader to the derivation in~\cite{belenchia}.}, is described by the same expression as in Equation~\eqref{eq:interference}, but with $\sin(\pi u)\rightarrow \pi u$ in Equation~\eqref{eq:talbot}. In order to claim the observation of quantum fringes, it is, thus, crucial to be able to distinguish the quantum fringe pattern from this \mbox{classical shadow.}}

\subsection{Accounting for decoherence and particle size}

The background theory presented above is a good foundation for understanding the quantum behaviour of our proposed experiment in the absence of decoherence. Nonetheless, in order to determine the viability of seeing interference fringes in the lab, we must account for all major sources of decoherence that will reduce the visibility of our fringes. We begin by describing the effects of decoherence events that have already been accounted for in previous works. These include collisions with residual gas particles, scattering and absorption of black-body photons, and thermal emission of radiation. The way these enter into our predicted interference pattern is by multiplying each Talbot coefficient, described in Equation~\eqref{eq:talbot}, by a reduction factor of the form:
\begin{equation}
R_n=\exp \left\{-\Gamma\left[1-f\left(\frac{n h t_2}{m D}\right)\right]\left(t_1+t_2\right)\right\},
\end{equation}
which washes out our expected interference fringes. $\Gamma$ here gives the rate of decoherence events, whilst $f(x)$ denotes their spatial resolution.

With these decoherence events accounted for, it is now time to consider the fact that our particle is not point-like, which is an assumption that has been implicitly made throughout the theory presented thus far. Relaxing this assumption has two major implications: it will change the form of the phase modulation parameter $\phi_0$ introduced in Equation~\eqref{eq:phi0} and impact the form of the reduction factors for scattering decoherence events that we are yet to account for. The point-like particle assumption is well justified in the Rayleigh scattering regime, that is when the radius of the particle is small compared to the wavelength of our optical grating, $kR << 1$, with $k$ being the standard wave number given by $k = 2 \pi / \lambda_G$. This will be mostly satisfied for the 
%simulation 
work presented here; however, our ambitions are to attempt performing this experiment for increasingly larger particles that are very close in size to the wavelength of the grating we intend to use. For a $10^8$~amu particle, $kR$ is already as high as $0.83$. This means we must work in the Mie scattering regime instead. In this regime, the form of the phase modulation parameter is adjusted to \cite{belenchia}
\begin{equation}
\phi_0 = \frac{8 F_0 E_G}{\hbar c \epsilon_0 a_G k |E_0|^2},
\end{equation}
where $E_0$ is the amplitude of our standing wave grating and $F_0$ is the force exerted on our particle, modelled by a dielectric sphere, by the grating according to Mie scattering theory. The expression for $F_0$ is long and not particularly enlightening; we, thus, refer the reader to~\cite{belenchia} for a detailed derivation and discussion. The key takeaways here are that the Rayleigh approximation predicts an increase in $\phi_0$ with the third power of the particle radius; this can be seen from Equation~\eqref{eq:phi0} since the real part of the particle's polarizability is proportional to the volume of the sphere. On the other hand, the Mie scattering correction predicts a steep sinusoidal fall-off in $\phi_0$ around the point $kR = 1$. Since this phase modulation parameter is exactly what ends up being converted into our spatial interference fringes, it is imperative that we avoid these regions of low $\phi_0$ when dealing with larger particles. A full graph of $F_0$ against $kR$ is presented in Figure 2 %MDPI: please confirm if it is this paper's Figure 2 citation, if yes, please revise it to Figure \ref{fig:nrgysVSxvT}. %No, this is the figure in reference [25]
 of \cite{belenchia}. Finally, we must also consider decoherence from the incoherent part of the scattering process and from the absorption of the grating photons. These fundamentally enter the expression for the interference pattern in the same way as the decoherence effects we have already discussed, but with more-complicated versions of $R_n$. For a full treatment of these terms, we once again refer to Belenchia et al. \cite{belenchia}; {see, however,~\cite{laing2023bayesian} and Appendix~\ref{appB} for a correction to Equation (28) of~\cite{belenchia}}. In the following, we include all these incoherent and coherent effects, treated in the full Mie scattering theory.
%, in our simulations. 
Note that we could also consider the misalignment of the grating relative to the particle's trajectory; however, this only leads to a net shift of the interference fringes and does not lead to a loss of visibility. This is discussed in the Supplementary Material of Bateman et al. \cite{bateman2014near} and will not be considered further here.

We now discuss a novel decoherence source specifically associated with the throw and catch scheme. This is the degree to which we are able to control the energy of the kicking laser pulse. An error in the kicking laser energy will cause an error in the flight time of the particle. If this error varies randomly over many runs that together build up our interference pattern, then this will lead to a smearing out of fringes as with other sources of decoherence. Unlike the other sources of decoherence mentioned earlier, this one is modelled by simulating many interference patterns, with normally distributed flight times that would be expected for an imperfect kicking laser, and then, averaged to create the final expected interference pattern. As with the requirement on cooling, we will see that the practical restriction of being able to recapture the particle is a stricter condition than what is needed to manage the decoherence effect coming from this error.

\section{Practical Considerations}

Firstly, it is important to mention the steps we expect to take to minimise the effects of decoherence to manageable levels. We minimise decoherence due to thermal emission of black body photons by minimising the rate of absorption, and, therefore, the rate of heating, of the particle whilst it is in the trap. Our silica particles have a particularly low absorption cross-section for the $1550$~nm light that we used to trap them. Decoherence by collisions with background gas particles is minimised by working at ultra-high vacuum ($10^{-10}$~mbar).
%Scattering and absorption decoherence at the grating laser is minimised by using a short ($5$~ns) pulse to limit the interaction time.

\begin{figure*}
\centering
\includegraphics[width=17.5cm]{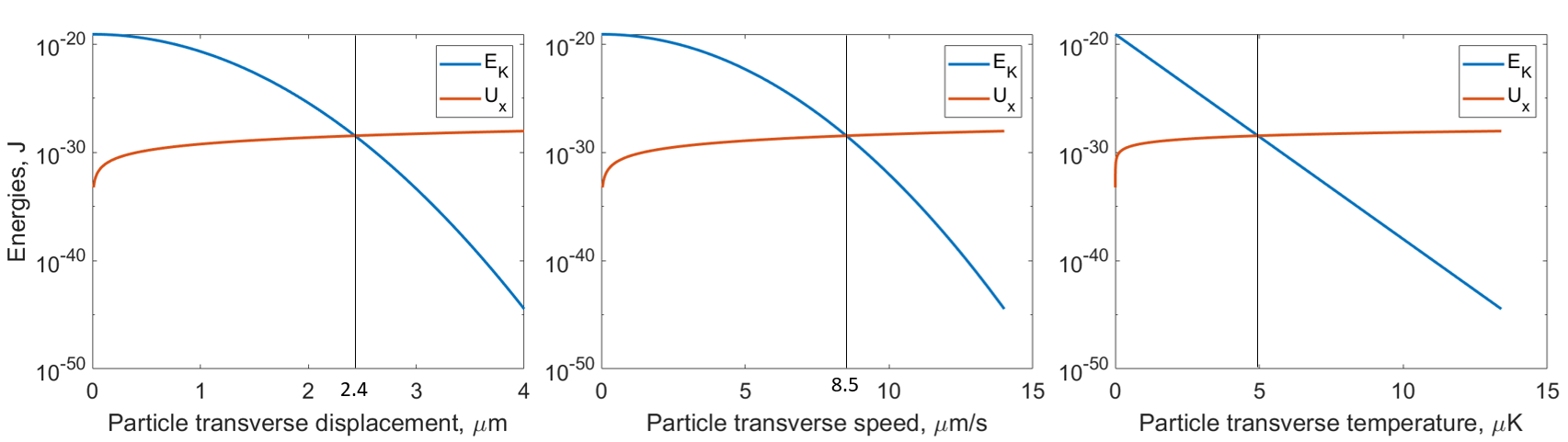}
	\caption{Matching %MDPI: 1.Please change the hyphen (-) into a minus sign (−, “U+2212”), e.g., “-1” should be “−1”; 2.there are no subfigure label in the figure body, please confirm; 3. this figure is not after first citation and over heading 1, please confirm. We've confirmed 2 and 3, we believe the minus signs are already minus signs. The chosen font-size might make it look otherwise but this was necessary for the overall clarity of the figure.
 point of optical potential energy barrier and particle kinetic energy.
	(\textbf{a}) The particle's transverse displacement must be less than 2.4~\textmu m for it to be recaptured.
	(\textbf{b}) To achieve that transverse displacement during the total flight time of $285.6$~ms, to and from a height of 10~cm, the transverse speed must be less than 8.5~\textmu m/s.
	(\textbf{c}) That speed corresponds to a temperature of the oscillation along $x$ of less than 5~\textmu K.}
	\label{fig:nrgysVSxvT}
\end{figure*}

With this in mind, we now discuss how practical restrictions on our ability to recapture the particle may end up being more restrictive on our potential interference pattern than any of the discussed decoherence events. Before we proceed, it is important to note that the strict cooling requirements that we discuss here could potentially be circumvented by tracking the particle and selecting to kick it only when it is in a low-velocity region of its oscillation. Whilst we will not discuss selection in more detail here, this is also an area of active investigation. In the following section, we will present the conditions needed to achieve an interferometer height of $10$~cm; this height is roughly what would be required to achieve flight times on the order of the Talbot time for a $10^6$~amu particle.
%\begin{itemize}
%\item Discuss the cooling we require for various heights.

We can begin to understand the cooling conditions on recapture by noticing that, when the particle is pushed vertically by the radiation pressure of the laser pulse, it will also have a transverse velocity $v_x$.
To recapture the returning particle, it must enter the optical trap at a transverse position where the optical potential barrier is greater than the transverse kinetic energy of the particle.
This transverse velocity is a measure of the particle's transverse temperature before release.
%The kinetic energy of the particle travelling at velocity $v_x$ is:
%\begin{equation}
%E_K = \frac{m v_x^2}{2}.
%\end{equation}
The optical potential energy barrier posed by the gradient force has the form \cite{hebestreit2017thermal}:
\begin{equation}
U_x = \frac{\operatorname{Re}(\alpha)}{2c\epsilon_0} I(x),
\end{equation}
where $\alpha$ is the electric polarizability of the particle and $I(x)$ is the intensity distribution of the laser along the $x$ direction.
We imposed the constraint that, when the particle reaches the focal plane ($z=0$), its kinetic energy is equal to that of the optical potential barrier.
The intensity distribution, assumed to be Gaussian, along $x$ is then:
\begin{equation}
    I(x,0,0) = \frac{2P}{\pi w_0^2} \exp\left[-2\left(\frac{x}{w_0}\right)^2\right],
\end{equation}
where $w_0$ is the laser waist in the focus of the parabola and $P$ is the power contained within the waist.
The potential energy as a function of $x$ becomes:
\begin{equation}
\begin{aligned}
U_x = \frac{\operatorname{Re}(\alpha) P }{c\epsilon_0 \pi w_0^2 }\exp\left[-2\left(\frac{x}{w_0}\right)^2\right].
\end{aligned}
\end{equation}
$x$ can also be written as a function of the (constant) radial velocity, $v_x$:
\begin{equation}
x = 2 v_x t = 2 v_x \cdot \sqrt{2 h / g},
\end{equation}
where $t = \sqrt{2 h / g}$ is the time needed for the particle to free fall from height $h$ and the \mbox{$2$ multiplication} comes from the fact that the particle travels sideways at $v_x$ during both the up and down motion along $z$.
The equality of the two energies becomes:
\begin{equation}
\frac{m v_x^2}{2} = \frac{\operatorname{Re}(\alpha) P }{c\epsilon_0 \pi w_0^2 }
\exp\left[-2\left(\frac{x}{w_0}\right)^2\right].
\end{equation}
%\footnote{Units check:
%$ m \cdot v^2 == \rm{kg \cdot m^2/s^2 = J } \\
%\alpha \cdot P \cdot \frac{1}{c} \cdot \frac{1}{\epsilon} \cdot \frac{1}{w^2} ==
%\rm{\frac{C^2 \cdot m}{N}\cdot \frac{N \cdot m}{s} \cdot \frac{s}{m} \cdot \frac{N %\cdot m^2}{C^2} \cdot \frac{1}{m^2} = N \cdot m = J}$.}
By graphically solving the above equation, we can then determine the cooling constraint on the maximal initial transverse displacement, initial transverse speed, and the corresponding temperature compatible with a throw and catch scheme of a 10~cm height. In Figure~\ref{fig:nrgysVSxvT}c, we can see that the temperature of the oscillator must be less than 5~\textmu K. The ground state temperature of our nanoparticle, when oscillating at $50$~kHz, is 2.2~\textmu K, and cooling down to this temperature has been achieved \cite{groundstatecooling}. This means that the cooling requirements for a $10$~cm height throw and catch are close to the state-of-the-art and, therefore, achievable. We also note that a lower frequency means larger displacement at the same temperature:
\begin{equation}
x_{rms} = \sqrt{\frac{k_B T}{m \omega_0^2}}
\end{equation}
Since particle displacement is what we are detecting, it may be easier to detect this more-ample motion at a lower frequency than a less ample motion at a higher frequency. A lower oscillation frequency would require less power, which reduces the heating and re-emission rate of the particle. These have been pointed out as problematic in \cite{bateman2014near}.

% \iffalse
% \begin{figure}[!htb]
% 	\centering
% 	\includegraphics[width=1\textwidth]{gstDisp.png}\\
% 	\caption{\title{Oscillator ground state temperature and displacement as a function of frequency.}
% 	\textbf{(a)} Oscillator ground state temperature as a function of frequency.
% 	 As the frequency of the oscillator increases, so does the temperature of its ground state.\\
% 	 \textbf{(b)} At a temperature of 9$mu$K, the displacement is greater at a lower oscillation frequency.}
% 	\label{fig:gstDisp}
% \end{figure}
% \fi

\begin{figure}[b]
\centering
\includegraphics[width=8.5cm]{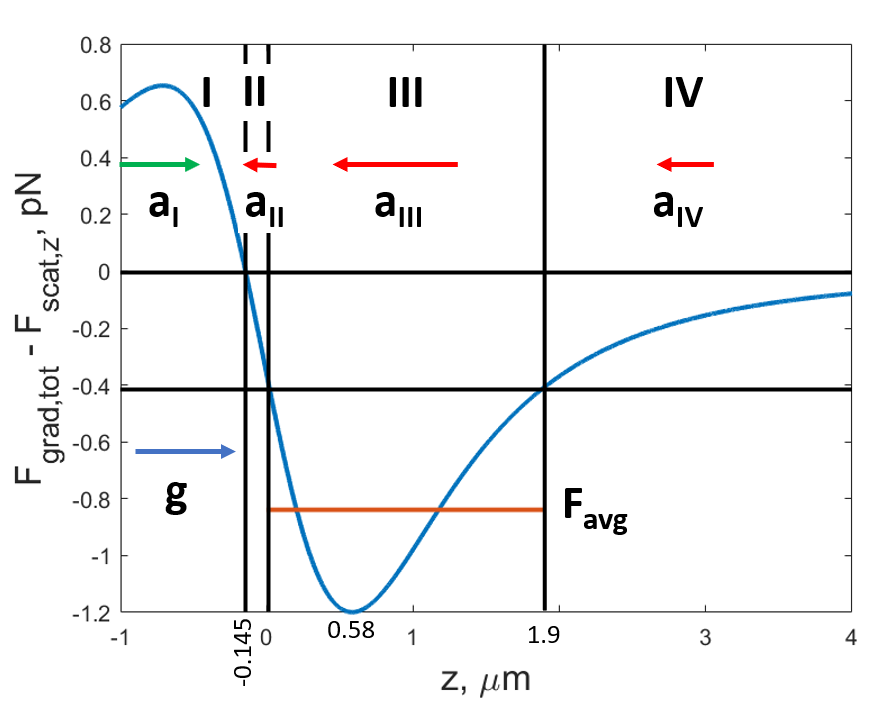}\\
\caption{Gradient %MDPI: 1.Please change the hyphen (-) into a minus sign (−, “U+2212”), e.g., “-1” should be “−1”. 2.please confirm if keep bold in figure caption.
 and scattering forces' equilibrium on the $z$-axis. The particle is falling from the left, in the direction of the gravitational acceleration, \textbf{g}, indicated by the blue arrow. Its diameter is $100$~nm, and the laser parameters are: $100$~mW, $1550$~nm. The focusing element is an NA $=$ 1 parabola of $3.6$~mm diameter. In the first region, \textbf{I}, where $F_{grad} > F_{scat}$, the particle receives an additional acceleration, $a_I$, from the gradient force. The second region, \textbf{II}, begins at $145$ nm in front of the focal plane, where $F_{grad} = F_{scat}$ for the particle considered, and ends at $z = 0$, where $F_{grad}=0$. Here, the deceleration begins, because the scattering force is greater than, and of opposite sign to, the gradient force. 
In the third region, \textbf{III}, both forces act in the direction opposite that of the particle. The deceleration, $a_{III}$, is highest here, with a peak at 0.58~\textmu m. 
The red, horizontal line, marks the average force of this region, denoted as $F_{avg}$. Its value is $-0.84$~pN. 
In the fourth region, after 1.9~\textmu m, the deceleration drops because both forces fade away as we depart from the focal plane. }
\label{fig:FgtmFsz}
\end{figure}

Now, turning our attention to the $z$ direction, the following can be said. If the particle is launched vertically at around $1$~ms$^{-1}$, then we can expect the same speed as it returns. The challenge now is to restart the laser at the right time in order to catch and re-cool the particle. The trapping of a particle can be understood as the balancing of two forces, a gradient force and a scattering force. The laser's electric field polarises the particle, and the gradient force pulls the dipole towards the region where the electric field is highest \footnote{This is true for a particle optically levitated in vacuum. If the refractive index of the medium is greater than that of the particle, then the particle is pushed away from the maximum field strength region.}, which is the centre of the focal spot. The scattering force arises due to the presence of the particle in the light field, which modifies the latter's energy flow \cite{rohrbach2001optical}. This force pushes the particle away from the centre of the trap. In Figure \ref{fig:FgtmFsz}, we show that a good time to re-start the trapping laser is when the particle is in region III. There, the gradient and scattering forces act in the direction opposite the direction of travel of the particle, and their contributions add to the greatest achievable deceleration. This region gives us approximately 2~\textmu m to stop the particle.
Region II can be thought of as the acceptable uncertainty in the arrival of the pulse that triggers the laser to turn back on.
The particle falling at around $1$~ms$^{-1}$ travels the $145$~nm of region II in $145$~ns.
If the laser is turned on at $t_0 \pm 145$~ns, where $t_0$ is the time at which the particle passes trough the focal plane ($z=0$), then the particle will be decelerated in the fastest time. Region I should be avoided since, in this region, the gradient force further accelerates the particle in the direction of gravity. We see that we now also have a relatively strict timing requirement to recapture the particle, as well as a strict cooling requirement. If the trap is turned on too soon, the gradient force will accelerate the particle too much for recapture to be possible, and if the trap is turned on too late, the gradient force will be too weak to stop the particle.

\begin{figure}[h]
	\centering
	\includegraphics[width=8.5cm]{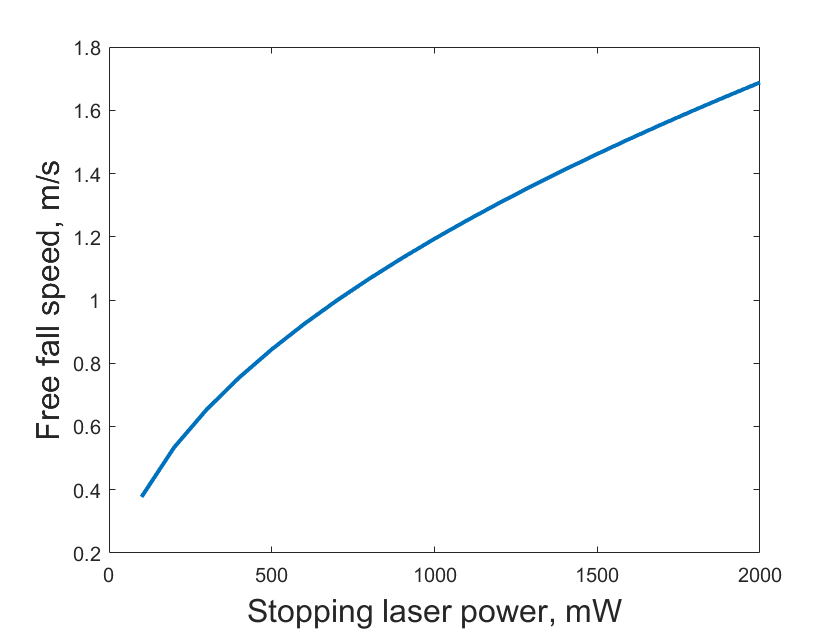}\\
	\caption{Particle return speed against trap power needed to stop it. To stop a particle returning at $1.4$~m/s, we need a laser power of $1.5$~W. }
	\label{fig:spdvspwr}
\end{figure}

The problem now is whether we can stop the particle in the available 2 \textmu m interval.
The stopping strength of the average force, $F_{avg}$, may be written as:
\begin{equation}
F_{avg}\cdot d = \frac{m v^2}{2},
\end{equation}
where $d$ is the distance over which the average force acts, $m$ is the particle mass, and $v$ is the highest velocity that the force can stop over this distance.
From here, we extract the maximal velocity to be:
\begin{equation}
v_{max} = \sqrt{\frac{2F_{avg}d}{m}} = \sqrt{\frac{2\cdot 8.4\cdot 10^{-13} \cdot 1.9 \cdot 10^{-6} }{2.2 \cdot 10^{-17}}} = 0.38~\rm{ms} ^{-1}.
\end{equation}
This is less than half of the expected return velocity. Figure \ref{fig:spdvspwr} shows the required laser power to stop a particle depending on its return velocity. We see that, in order to stop, for example, a $1.4$~ms$^{-1}$ particle, we would need a laser power of $1.5$~W, whilst our trap typically operates at $100$~mW. We must note that only the scattering force is dissipative and actually takes away energy from the falling particle.
The gradient force will spring the particle back upwards after the latter was brought to a halt. 
So, in order to properly brake the particle, one would need to modulate the laser power, in synchronisation with the motion of the particle, starting from $1.5$~W down to the $100$~mW stationary trapping power, until the latter is brought to the stationary regime, where the regular feedback cooling protocol can be applied.

\section{Results} 

\subsection{Expected Interference Patterns}

\begin{figure*}
\centering
\includegraphics[width=15.5cm]{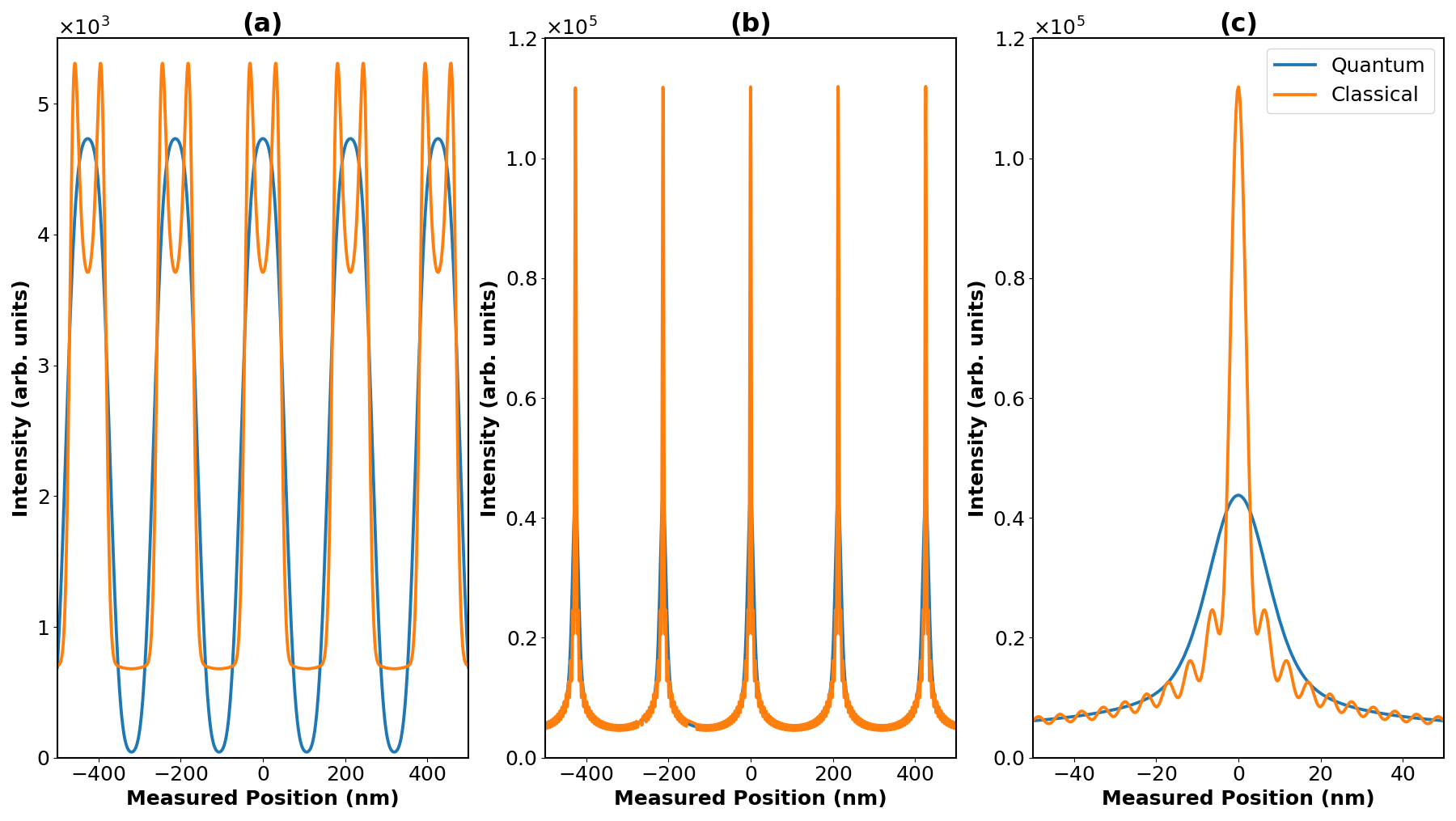}
\caption{Expected %MDPI: Please change the terms into scientific notations in the figure, e.g., “8 × 10−4”, not “8e3”.
 quantum interference patterns (blue) and Moiré shadow patterns (orange) for a (\textbf{a}) $10^6$ amu nanoparticle and (\textbf{b}) a $10^8$ amu nanoparticle. (\textbf{c}) shows a zoomed in section of (\textbf{b}) for clarity. These patterns assume parameters that were found by optimising for the visibility of the quantum fringes in each case and are detailed in Table~\ref{tab1}. These patterns include all interference effects mentioned in the previous section, aside from the decoherence from imperfect kicks, which is shown later. \label{fig2}}
\end{figure*}

We now present our expected interference patterns for particles of mass $10^6$ and $10^8$ atomic mass units, as well as how the fringe visibility varies as we change various parameters. Figure~\ref{fig2} shows both the expected quantum interference patterns, with decoherence sources included, and the classically predicted Moiré shadow patterns for the two masses of particles we investigated. The parameters we assumed are shown in Table~\ref{tab1} \footnote{Despite the cooling requirements detailed in the previous section, we chose a 1 mK temperature here to demonstrate the relatively low cooling requirements needed to see visible fringes. In theory, should a better solution for recapture be found, this would be the new cooling requirement. Using the temperatures from the previous section would lead to slightly higher visibility fringes.}.

\begin{table}[h] 
\caption{Table of relevant parameters used for the generation of quantum and classical patterns throughout this section.\label{tab1}}
\begin{ruledtabular}
\begin{tabular}{ccc}
\textrm{\textbf{Parameter}}&
\textrm{$10^6$ amu particle}&
\textrm{$10^8$ amu particle}\\
\colrule
Initial temperature		& 1mK			& 1mK\\
Pressure		& $10^{-10}$ mbar			& $10^{-10}$ mbar\\
Flight time		& 58 ms			& 142 ms\\
Phase modulation($\phi_0$)		& $\pi / 2$			& $8\pi$\\
\end{tabular}
\end{ruledtabular}
\end{table}

An important point to make about the parameter selection shown in Table~\ref{tab1} is the short flight time selected for the high-mass particle. We mentioned previously that, to be confident about seeing quantum interference effects, the state should be allowed to evolve for a time on the order of the Talbot time before interacting with the grating. The Talbot time for the $10^8$~amu particle is around $2.8$~s, and yet, we selected a time of 142~ms, an order of magnitude lower. The reasoning for this is that, in order to achieve a flight time on the order of the Talbot time, the interferometer would have to be much taller, on the order of meters or tens of meters. We would still need to recapture the particle in our optical trap, which has a recapturing range on the order of microns. For recapture to be achieved for these flight times, we would require levels of cooling several of orders of magnitude beyond what has been achieved experimentally to date. For this reason, we restricted our investigation to tens or hundreds of milliseconds, resulting in significantly sub-Talbot times for the higher-mass case. Nonetheless, our numerical estimates
%simulations 
showed clear high-visibility fringes even when working in this regime, as demonstrated in Figure~\ref{fig2}.

\subsection{Quantum/Classical Distinctions}

\begin{figure*}
\centering
\includegraphics[width=14cm]{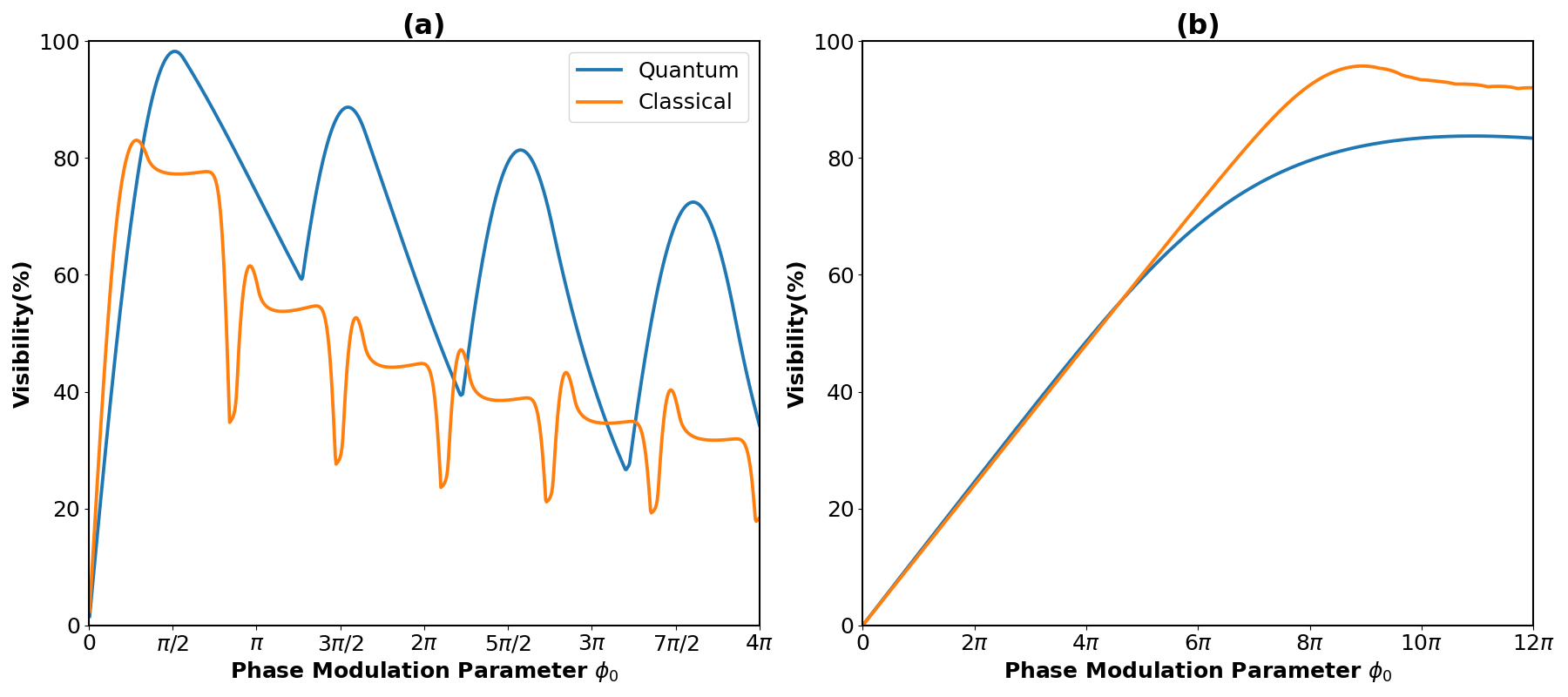}
\caption{Expected variation of quantum (blue) and classically (orange) predicted visibilities of fringes when varying the phase modulation parameter $\phi_0$, controlled by the pulse energy of the grating laser. (\textbf{a}) shows the expected visibility variation for a SiO\textsubscript{2} particle of mass $10^6$~amu, whilst \mbox{(\textbf{b}) demonstrates} the same plot for a particle of mass $10^8$~amu.\label{fig3}}
\end{figure*}

\begin{figure*}
\centering
\includegraphics[width=14cm]{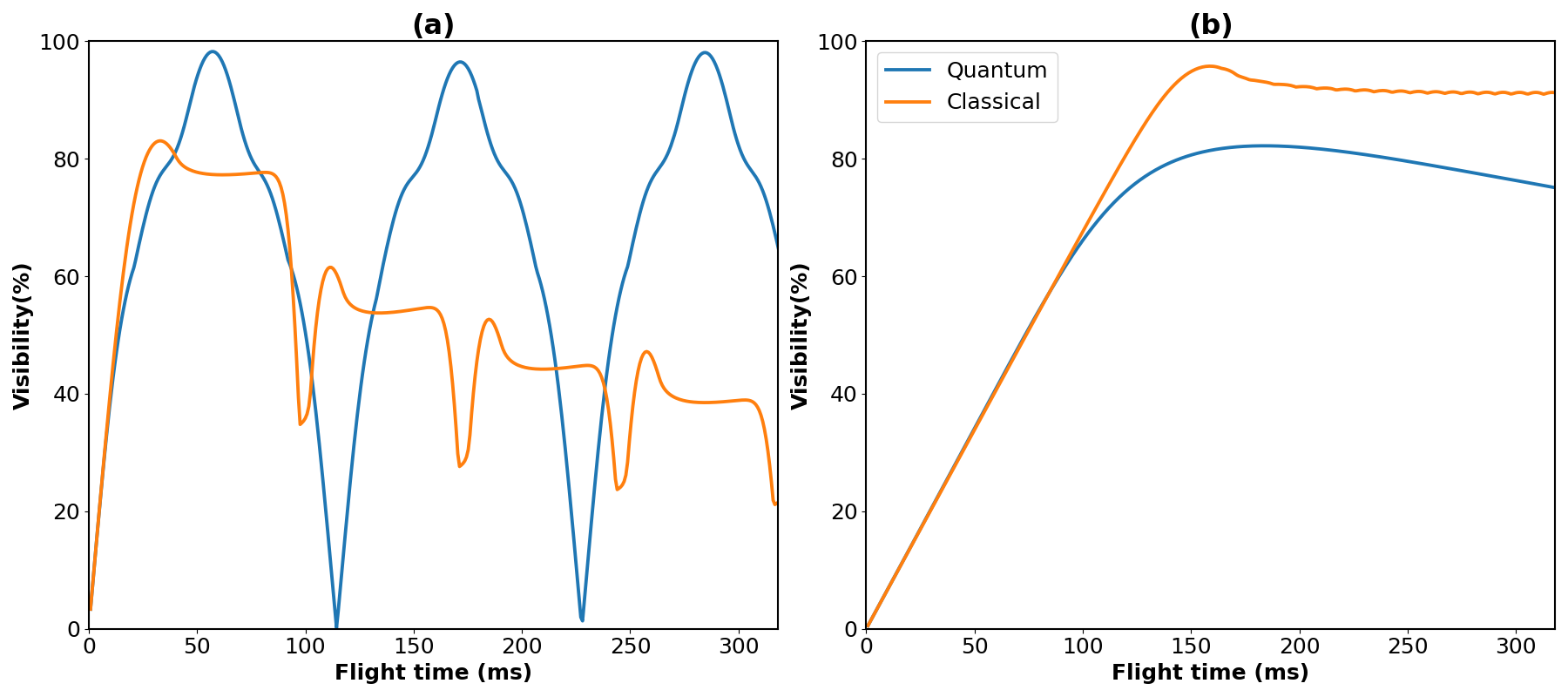}
\caption{Expected variation of quantum (blue) and classically (orange) predicted visibilities of fringes with varying flight times. (\textbf{a}) shows visibility variation for a $10^6$~amu SiO\textsubscript{2} particle, whilst \mbox{(\textbf{b}) shows} the same plot for a $10^8$~amu particle. \label{fig4}}
\end{figure*}

As we see in Figure~\ref{fig2}b, it can be quite difficult to distinguish between the quantum and classical predictions, especially when working with the higher-mass particles in the sub-Talbot regime. Distinguishing between the two predictions in the high-mass case would require spatial resolution in our measurements on the order of nanometres. While not impossible, this would certainly be another highly demanding requirement. For this reason, we propose alternative means by which to differentiate between the quantum and classical case. By generating many interference patterns whilst varying either the phase modulation parameter ($\phi_0$) or the flight time and recording the visibility of the observed fringes, we can construct plots with clearer distinctions for the quantum and classical case. \mbox{Figures~\ref{fig3} and~\ref{fig4}} demonstrate this idea. Aside from the ones that are changing, the parameters used for these plots are the same as those detailed in Table~\ref{tab1}.

The differences in the quantum and classical predictions was once again most clear for the $10^6$~amu particle, which has the privilege of working in the Talbot regime. Nonetheless, we now see distinctive differences in the predicted visibilities of the quantum and classical models even for the $10^8$~amu particle. The models diverged from each other when dealing with a longer time of flight and larger values of the phase modulation parameter. By conducting such experiments, the quantum and classical descriptions could be more easily distinguished when the required spatial resolution to distinguish them from a single interference pattern is too challenging to achieve. A final point to make is that the parameters used in our numerical analysis
%within these simulations 
were chosen by optimising for the highest visibility of quantum fringes in each case. It would be entirely possible to instead optimise for the largest divergence between quantum and classical visibility predictions instead, whilst sacrificing overall visibility. Since our predicted visibilities are relatively high, even when accounting for major sources of decoherence, this could be a valuable method to ensure our ability to distinguish between quantum and classical predictions.

\begin{figure*}
\centering
\includegraphics[width=15cm]{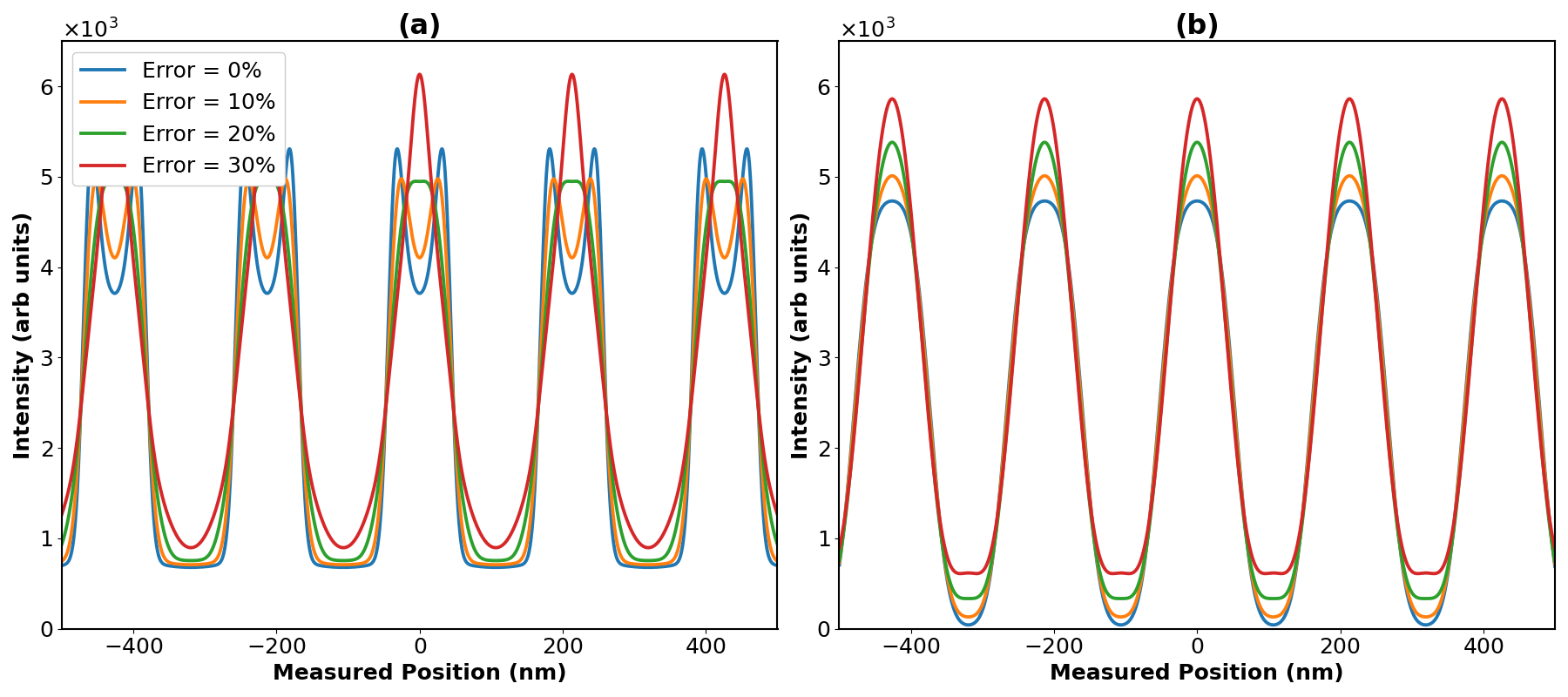}
\caption{Expected %MDPI: Please change the terms into scientific notations in the figure, e.g., “8 × 10−4”, not “8e3”.
 interference patterns for the interferometry of a $10^6$~amu SiO\textsubscript{2} particle when accounting for the error in initial kick velocity, with errors ranging from 0 to 30\%. (\textbf{a}) shows the expected patterns for the classical case, whilst (\textbf{b}) shows the quantum case. \label{fig6}}
\end{figure*}  

\begin{figure*}
\centering
\includegraphics[width=15cm]{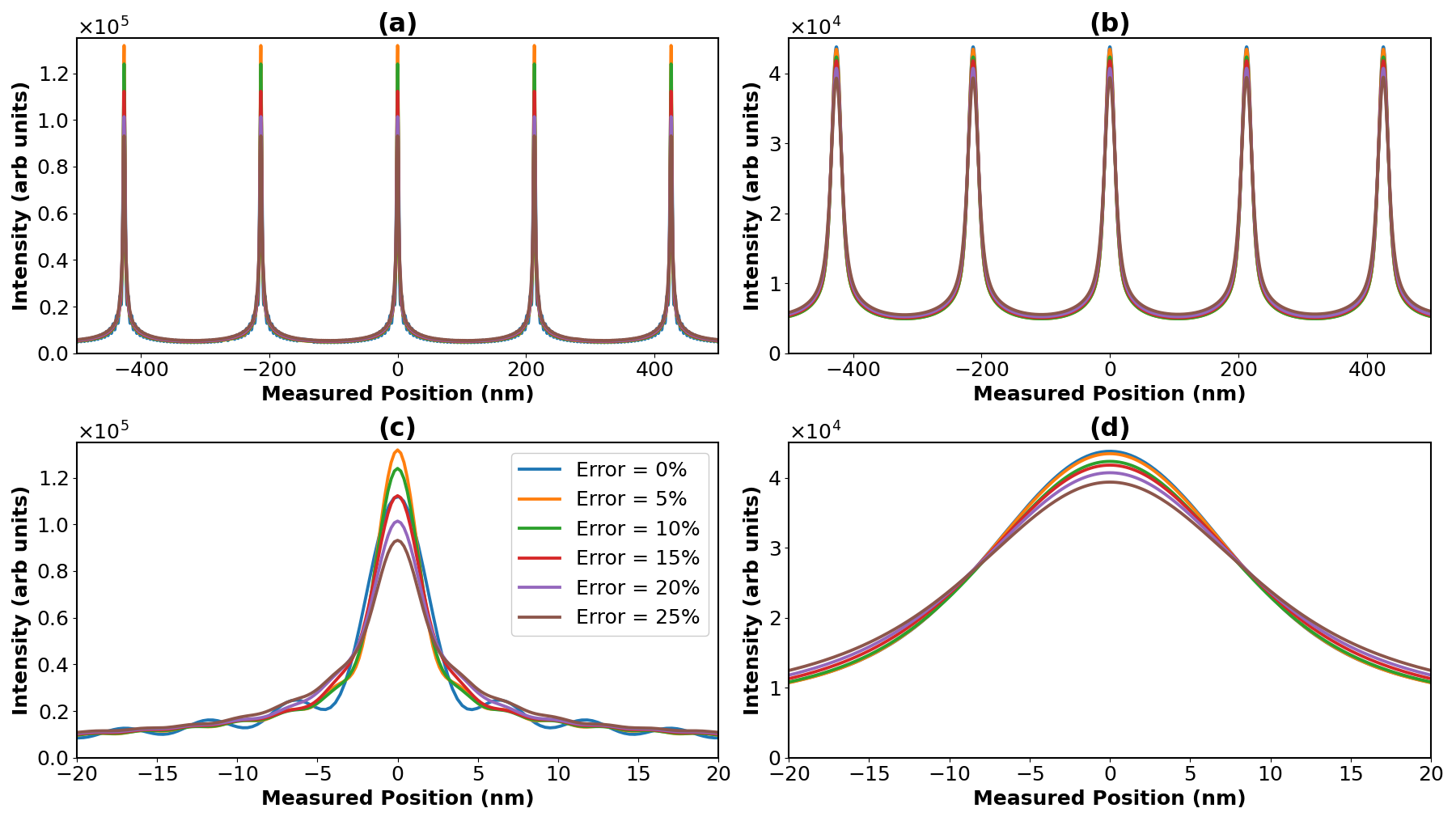}
\caption{Expected %MDPI: Please change the terms into scientific notations in the figure, e.g., “8 × 10−4”, not “8e3”.
 interference patterns for the interferometry of a $10^8$~amu SiO\textsubscript{2} particle when accounting for the error in the initial kick velocity, with errors ranging from 0 to 25\%. (\textbf{a}) shows the expected patterns for the classical case, whilst (\textbf{b}) shows the quantum case. (\textbf{c},\textbf{d}) give a zoomed in picture of (\textbf{a},\textbf{b}), respectively.  \label{fig5}}
\end{figure*}  

\newpage

\subsection{Throw and Catch Specific Decoherence}

As previously mentioned, the plots shown up to this point do not consider the decoherence associated with imperfect control of the energy of the kicking pulse and, therefore, imperfect flight times. We now demonstrate the effects of this source of decoherence on the visibility of our fringes. Figures~\ref{fig6} and~\ref{fig5} show how varying errors of initial velocity impact the visibility for the case of a $10^6$~amu particle and a $10^8$~amu particle, respectively. Once again, the parameters for these plots are the same as those detailed in Table~\ref{tab1}. These plots were generated by simulating many interference patterns, with flight times varying according to a normally distributed velocity profile for the initial kick, and then, averaged into the overall expected pattern. In these plots, the error refers to the standard deviation of this velocity distribution.

The results shown in Figure~\ref{fig5} are not particularly surprising since we are working in the sub-Talbot regime. As we saw in Figure~\ref{fig4}b, this means that variation in the visibility with the flight time is relatively slow and smooth. Hence, we would expect a slow decrease in visibility with an increase in the initial velocity error, whilst not expecting a significant change in the shape of the pattern. This is exactly what Figure~\ref{fig5} shows, although we also see the washing out of the detail beyond the central fringe for the classical case. Figure~\ref{fig6}, on the other hand, is a bit different. Whilst the quantum case once again shows a slow decrease in visibility, the classical case now also shows a change in the shape of the pattern. In fact, we see that the overall positioning of the fringes for the two regimes converges with increasing error. This could mean that the quantum and classical patterns could look quite similar to each other if the error in the initial kick velocity was high, even for a $10^6$~amu particle. Nonetheless, the key takeaway from these plots is that even very large errors of 25\% or 30\% still led to high-visibility fringes, which implies this is a very manageable source of extra decoherence. The explicit visibility values for the plots are provided below in Tables~\ref{tab2} and~\ref{tab3}.

\begin{table}[h] 
\caption{Table of visibility values for the plots presented in figure~\ref{fig6} for a $10^6$~amu particle.\label{tab2}}
\begin{ruledtabular}
\begin{tabular}{ccc}
\textbf{Error}&\textbf{Classical Visibility}&\textbf{Quantum Visibility}\\
\hline
0\%		& 77.3\%	& 98.2\%\\
10\%	& 75.0\%	& 94.9\%\\
20\%	& 73.5\%	& 88.3\%\\
30\%	& 74.5\%	& 81.1\%\\
\end{tabular}
\end{ruledtabular}
\end{table}

\begin{table}[h] 
\caption{Table of visibility values for the plots presented in figure~\ref{fig5} for a $10^8$~amu particle.\label{tab3}}
\begin{ruledtabular}
\begin{tabular}{ccc}
\textbf{Error}&\textbf{Classical Visibility}&\textbf{Quantum Visibility}\\
\hline
0\%		& 92.2\%	& 79.6\%\\
5\%		& 93.0\%	& 79.6\%\\
10\%	& 92.5\%	& 79.1\%\\
15\%	& 91.5\%	& 78.4\%\\
20\%    & 90.3\%	& 77.2\%\\
25\%    & 89.0\%	& 75.5\%\\
\end{tabular}
\end{ruledtabular}
\end{table}

\subsection{Experimental Progress}

\begin{figure*}
\includegraphics[width=15cm]{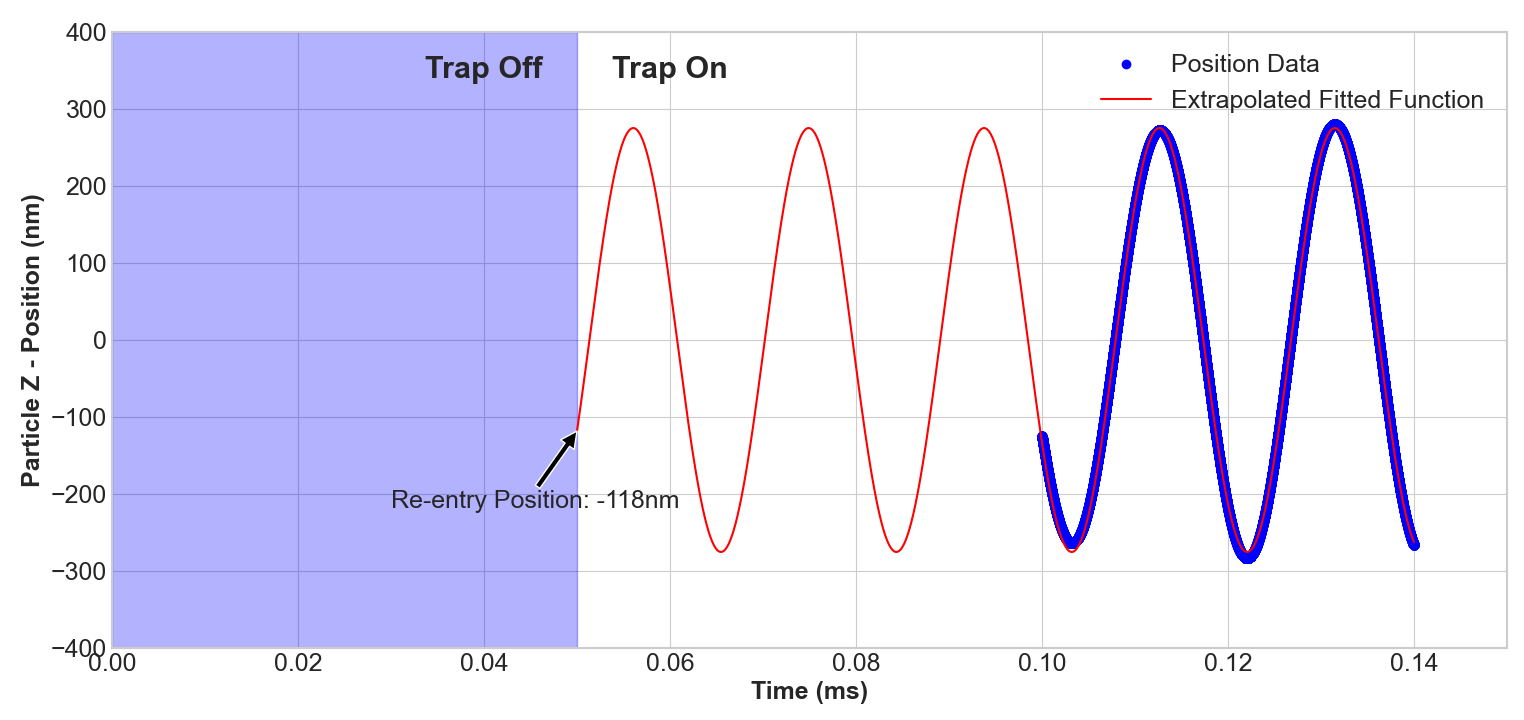}
\centering
\caption{Example of a measurement of a particle's re-entry position in the vertical ($z$) direction. The trap was turned off at $0$~ms and turned back on at $0.05$~ms. Position is measured from the equilibrium position of the particle, which is slightly offset from the centre of the trap. \label{fig7}}
\end{figure*}  

Since much has already been published on the trapping and cooling aspects that will be used in this experiment, here, we give a preliminary demonstration of how the measurement portion of the interferometer will be achieved. Inspired by \mbox{Hebestreit et al. \cite{measurementconcept},} we tested our measurement scheme by performing release and recapture experiments. Once the particle was prepared in the state we desired before starting interferometry, we turned off the trap for 50~\textmu s and, then, turned it back on. We then observed the oscillations immediately after the trap was turned on. By filtering for a desired direction of oscillation and then fitting to the filtered data, we were able to extract the re-entry position. This procedure is illustrated in Figure~\ref{fig7}.

We can estimate a rough precision of this technique by taking different-length segments of the data, applying the same process, and seeing how much the re-entry position varied as a result. When this was performed, we found that all extracted positions lied within $15$~nm of each other. This should already be sufficient for our purposes. This was just a demonstration of the idea behind our detection scheme and not the final product. Work is ongoing for using more-sophisticated techniques for extracting the position at which the particle re-enters the trap, such as using a particle filter \cite{particlefilter}. These techniques could also potentially allow us to implement more-consistent recapture methods without significantly affecting the accuracy of the re-entry position measurement. An example of one such method would be to re-apply the kicking laser to slow down the particle prior to it re-entering the trap. This could alleviate the strict requirements we currently have on the timing of when we turn on the trap or, equivalently, on the initial launch velocity.

\newpage

\section{Conclusion} 
By building upon previous ideas for how wave-like behaviour could be observed for a high-mass particle, we presented a throw and catch scheme that we believe will be able to generate quantum interference patterns of nanoparticles, with masses up to three order of magnitude greater than the current matter wave interferometry record. Whilst maintaining the easy-to-achieve conditions of previous proposals, such as a room temperature environment and using only a single optical diffraction element, we showed how the problems associated with inefficient loading can be sidestepped without the introduction of significant extra decoherence. We showed how the Talbot effect can be used to generate similar interference patterns to what has been discussed previously with our new proposed setup. Accounting for all major sources of decoherence and working with similar masses to what has been discussed before, that being $10^6$~amu particles, we found the possibility for high-visibility fringes with a clear distinction between the quantum and classical predictions, all while adhering to the practical limitations associated with recapturing the particle. We showed that this system could even be used for particle masses in the region of $10^8$~amu by working with flight times below the Talbot time, although, in this case, with more similarity between the quantum and classical predictions. 

\begin{acknowledgments}
We thank Sarah Waddington for discussions. We would like to thank the reviewers for their insightful comments on re-applying the kicking laser as an alternate method for recapture. We acknowledge the use of the IRIDIS High Performance Computing Facility, and associated support services at the University of Southampton, in the completion of this work. We acknowledge support from the QuantERA grant LEMAQUME consortium, funded by the QuantERA II ERA-NET Cofund in Quantum Technologies implemented within the EU Horizon 2020 Programme. Further, we would like to thank the UK funding agency for the funding from EPSRC under grants EP/W007444/1, EP/V035975/1, EP/V000624/1, and EP/X009491/1, the Leverhulme Trust (RPG-2022-57), the EU Horizon 2020 FET-Open project {\it TeQ} (GA No. 766900), and the EU Horizon Europe EIC Pathfinder project {\it QuCoM} (GA No. 10032223).
{
G.G. also acknowledges the support from QuantERA grant C’MON-QSENS!, by Spanish MICINN PCI2019-111869-2,
the Spanish Agencia Estatal de Investigación, project PID2019-107609GB-I00/AEI/10.13039/501100011033,
the project Spanish MCIN (project PID2022-141283NB-I00) with the support of FEDER funds, by the Spanish MCIN with funding from European Union NextGenerationEU (grant PRTR-C17.I1), and the Generalitat de Catalunya, as well as the Ministry of Economic Affairs and Digital Transformation of the Spanish Government through the QUANTUM ENIA `Quantum Spain' project with funds from the European Union through the Recovery, Transformation and Resilience Plan - NextGenerationEU within the framework of the `Digital Spain 2026 Agenda'.
}
\end{acknowledgments}

\bibliography{biblio}% Produces the bibliography via BibTeX.

\setcounter{figure}{-1}
\renewcommand*{\thefigure}{A\arabic{figure}}
\begin{figure*}
\centering
\includegraphics[width=15cm]{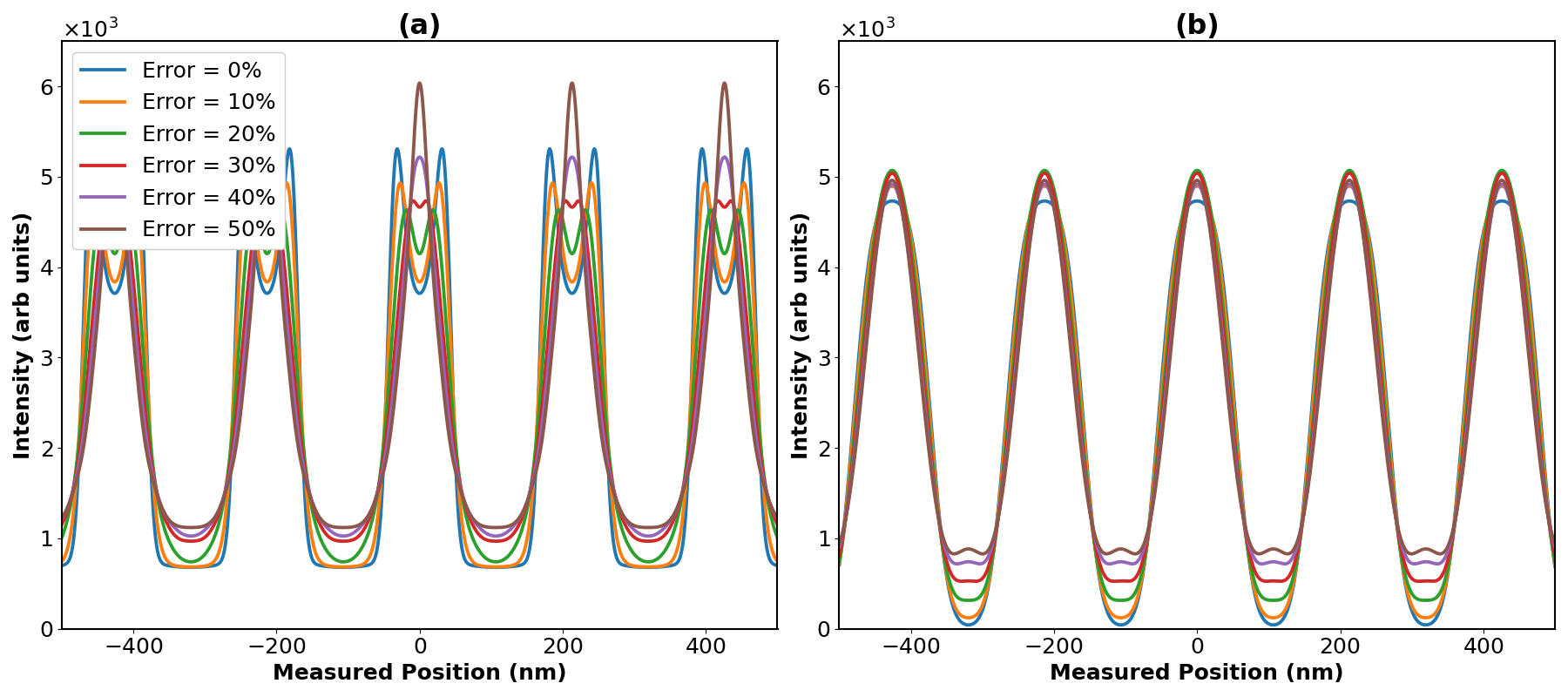}
\caption{Expected  %MDPI: Please change the terms into scientific notations in the figure, e.g., “8 × 10−4”, not “8e3”.
 interference patterns for the interferometry of a SiO\textsubscript{2} particle, with a mean mass of $10^6$~amu, when accounting for a spread of different particle masses being used for each run of the experiment. Errors in mass range from 0 to 50\%. (\textbf{a}) shows the expected patterns for the classical case, whilst (\textbf{b}) shows the quantum case. \label{fig:A1}}
\end{figure*}  
\newpage
\appendix

\section{Incoherent sources}

\begin{table}[ht]
\caption{Table of visibility values for the plots presented in figure~\ref{fig:A1}. \label{tabA1}}
\begin{ruledtabular}
\begin{tabular}{ccc}
\textbf{Mass Error}&\textbf{Classical Visibility}&\textbf{Quantum Visibility}\\
\hline
0\%		& 77.3\%	& 98.2\%\\
10\%	& 75.6\%	& 95.1\%\\
20\%	& 72.4\%	& 88.3\%\\
30\%	& 66.0\%	& 81.1\%\\
40\%    & 67.1\%	& 74.5\%\\
50\%    & 68.7\%	& 71.4\%\\
\end{tabular}
\end{ruledtabular}
\end{table}
To demonstrate the advantage of re-using the same SiO$_2$ particle for every run of our interferometer, we performed simulations that allowed the mass of the particle used for each run to vary. This would roughly simulate the situation where a new particle would be reloaded for each run. Figure~\ref{fig:A1} shows the results of these simulations when particles of a mean mass of $10^6$~amu were used, with other parameters matching those listed in Table~\ref{tab1}. Table~\ref{tabA1} shows that the visibility of the fringes, that we expect to see in the quantum interference pattern, decreased as the spread of masses that the particles can take increased. In Figure~\ref{fig:A1}, we also see that, if the spread in the mass of the particles exceeded \mbox{$30\%$, then} the maximal intensity regions of the quantum and classically predicted interference patterns would overlap, leading to additional difficulty in distinguishing between the \mbox{two cases.} Therefore, we see that re-using the same particle can lead to drastic increases in visibility depending on the variation in mass of the different particles that would be used otherwise. It is also important to note that these simulations only took into account a change in mass of the particle. In reality, different particles could have different shapes or densities depending on the purity of the batch that is being used. This would lead to even further washing out of the fringes. We also note that these visibility drops could be partially avoided, when not re-using the same particle, by post-selecting data where only similar particles were used. However, this would lead to more runs being necessary, where, again, the problem of inefficient re-loading arises.

\begin{widetext}

\section{Mie Scattering Correction}\label{appB}

As stated in the main text, in our numerical analysis, we implemented a correction with respect to the results in~\cite{belenchia} as concerns the decoherence reduction factor induced by the scattering of grating photons in the Mie scattering regime, i.e., accounting for the finite size of the particles.

We closely followed the treatment in~\cite{belenchia}. Under the assumption that the laser waist is  much larger than the size of the particle, the effect of the scattering of the grating's photons is described by the action of the Lindblad super-operator:  %MDPI: please confirm if keep all bold of equations in the manuscript.

\begin{equation}
    \mathcal{L}[\rho_S] = |\alpha|^2 \sum_{\nu}\int d\bm{k}\delta(\omega_{k}-\omega_0)\left(2\mathcal{T}_{\bm{k}\nu,c}(\hat{r})\rho_S\mathcal{T}^*_{c,\bm{k}\nu}(\hat{r})-\left\{|\mathcal{T}_{\bm{k}\nu,c}(\hat{r})|^2,\rho_S\right\}\right)%\nonumber\\
\end{equation}
where the collisional operators are 
\begin{align}
\mathcal{T}_{\bm{k}\nu,\bm{c}}(\hat{\bm{r}})= \sum_{\nu'}\int d\bm{k}' \langle{\bm{c}}|{\bm{k}',\nu'}\rangle\mathcal{T}^{*}_{\bm{k}'\nu',\bm{k}\nu}(\hat{\bm{r}}).
\end{align}
For our case of interest, $\ket{\bm{c}}$ is a linearly polarised standing wave with mode volume $V_0$ such that
\begin{equation}
\begin{array}{l}
\bra{\bm{k},\nu}{\bm{c}}\rangle  = \frac{1}{V_{0}}\int d\bm{x}\, e^{-i\bm{k}\cdot \bm{x}} \bm{\epsilon}_{\bm{k},\nu}\cdot \bm{\epsilon_{d}}\,g(x,y)\cos(k_{0}z)\\
~~~~~~~~~~~~~\simeq \frac{\bm{\epsilon}_{\bm{k},\nu}\cdot\bm{\epsilon}_{k_{z},\nu'}}{\sqrt{V_{0}}}\tilde{g}(k_{x},k_{y})\delta(k_{z}^{2}-\omega_{0}^{2}) \simeq \frac{\omega_{0}\bm{\epsilon}_{k,\nu}\cdot\bm{\epsilon}_{k_{z},\nu'}}{\sqrt{V_{0}}}\tilde{g}(k_{x},k_{y})\delta(k_{x})\delta(k_{y})\delta(k_{z}-\omega_{0}^{2}).
\end{array}
\end{equation}
where g(x,y) represents the laser transverse beam profile and $\tilde{g}(k_{x},k_{y})$ its Fourier transform.
Notice that the approximation made in the last line is justified under the assumption of a laser field with a very wide spot area $a_{g}$, i.e., $a_{g}\gg k_{0}^{2}$

We, thus, have, for the collisional operators, 
\begin{align}
\mathcal{T}_{\bm{k}\nu,\bm{c}}(\hat{\bm{r}})\simeq \frac{1}{4\pi \omega_{k}\sqrt{V_{0}}} \left(e^{-i\bm{k}_{0}\cdot \hat{\bm{r}}}f^{*}_{\nu_{0},\nu}(\bm{k_{0}},\bm{k})+e^{i\bm{k}_{0}\cdot \hat{\bm{r}}}f^{*}_{\nu_{0},\nu}(-\bm{k_{0}},\bm{k}) \right),
\end{align}
with $f _{\nu,\nu'}(\bm{k},\bm{k}')$ the Mie scattering amplitude.

Considering the incoming standing wave polarised along the $x$ direction, we obtain the vectorial scattering amplitude: 
\begin{align}
\sum_{\nu} \epsilon_{\nu} f^{*}_{\nu,{x}}(\bm{k_{0}},\bm{k})=(S_{2}\cos\phi \epsilon_{\theta}-S_{1}\sin\phi \,\epsilon_{\phi}).
\end{align}
Here, $\epsilon_{\theta} \cdot \epsilon_{\phi}=0$, as they are two orthogonal components of the scattered field polarisation. For the explicit expressions for $S_1$ and $S_2$, we refer the reader to Appendix A %MDPI: please confirm if it this paper's Appendix citation, if yes, please change it to Appendix \ref{app1}. No, this is the appendix A of ref [25].
 of~\cite{belenchia}.

We, then, finally, arrive at the rewritten Liouvillian super-operator in the form:
\begin{equation}
\begin{array}{l}
    \mathcal{L}[\rho_S] = |\alpha|^2 \sum_{\nu}\int d\bm{k}\delta(\omega_{k}-\omega_0)\left(2\mathcal{T}_{\bm{k}\nu,c}(\hat{r})\rho_S\mathcal{T}^*_{c,\bm{k}\nu}(\hat{r})-\left\{|\mathcal{T}_{\bm{k}\nu,c}(\hat{r})|^2,\rho_S\right\}\right)\\
   \hspace{+27pt} = |\alpha^{2}| \int d\bm{k} \delta(\omega_{k}-\omega_{0})\left(2\mathcal{T}_{\bm{k}\phi,c}(\hat{r})\rho_S\mathcal{T}^*_{c,\bm{k}\phi}(\hat{r})-\left\{|\mathcal{T}_{\bm{k}\phi,c}(\hat{r})|^2,\rho_S\right\}\right)\\
   \hspace{+27pt} +|\alpha^{2}| \int d\bm{k} \delta(\omega_{k}-\omega_{0})\left(2\mathcal{T}_{\bm{k}\theta,c}(\hat{r})\rho_S\mathcal{T}^*_{c,\bm{k}\phi}(\hat{r})-\left\{|\mathcal{T}_{\bm{k}\theta,c}(\hat{r})|^2,\rho_S\right\}\right)
    \end{array}
\end{equation}
From this last expression, following~\cite{belenchia}, we, finally, arrive at the correct form of their Equation~(28), which reads
\begin{align}
R_{\rm sca}(z_{-},z_{+})= \exp\left[F(s)+a(s)\cos (2 k z)+ib(s)\sin(2kz) \right],
\end{align}
with, now, the coefficients given by
\begin{align}\label{abF}
a(s) &= \frac{2\pi c}{V_0}\int d\tau|\alpha(\tau)|^{2}\int d \Omega\,{\rm Re}\big(\bold{f}^{*}(k , k \bold{n})\cdot\bold{f}(- k , k\bold{n})\big)[\cos(k n_{z} s)-\cos(ks)],\nonumber\\
b(s)&=  \frac{2\pi c}{V_0}\int d\tau|\alpha(\tau)|^{2}\int d \Omega \,{\rm Im} \big(\bold{f}^{*}(k , k \bold{n})\cdot \bold{f}(- k , k\bold{n})\big)\sin(kn_{z}s),
\nonumber\\
F(s)&=  \frac{2\pi c}{V_0}\int d\tau |\alpha(\tau)|^{2}\int d \Omega\, |\bold{f}(k,k\bold{n})|^{2} [\cos((1-n_{z})ks)-1].%\appendix
\end{align}
with $\mathbf{f}(k,k\bold{n}) =(f^{\parallel}(k,k\bold{n}),f^{\perp}(k,k\bold{n}))^{T}$ and
\begin{align}
f^{\parallel}(k,k\bold{n})=S_{1} \cos\phi\\
f^{\perp}(k,k\bold{n})=S_{2} \cos\phi
\end{align}
\end{widetext}

\end{document}